\documentclass[aps,prc,superscriptaddress,showpacs,onecolumn,preprint,showkeys]{revtex4}	
 \usepackage{amsmath}  
 \usepackage{makeidx}
 \usepackage{graphicx}
 \usepackage{amsfonts}
 \usepackage[ansinew]{inputenc}
 \usepackage{subfigure}
 \usepackage{color}
 \usepackage{mathrsfs}

\begin{document}
\title{Medium effects on freeze-out of light clusters at NICA energies}
\author{G.~R\"{o}pke}
\email{gerd.roepke@uni-rostock.de}
\affiliation {Institut f\"{u}r Physik, Universit\"{a}t Rostock,  Rostock, Germany}
\affiliation {National Research Nuclear University (MEPhI),  Moscow, Russia}
\author{D.~Blaschke }
\email{blaschke@ift.uni.wroc.pl}
\affiliation {National Research Nuclear University (MEPhI),  Moscow, Russia}
\affiliation {Joint Institute for Nuclear Research (JINR), Dubna, Russia}
\affiliation {Institute of Theoretical Physics, University Wroclaw, Poland}

\author{Yu.~B.~Ivanov}
\email{y.ivanov@gsi.de}
\affiliation {National Research Nuclear University (MEPhI),  Moscow, Russia}
\affiliation {Joint Institute for Nuclear Research (JINR), Dubna, Russia}
\affiliation {National Research Center "Kurchatov Institute",  Moscow, Russia}

\author{Iu.~Karpenko }
\email{yu.karpenko@gmail.com}
\affiliation {SUBATECH, Universit\'e de Nantes, Nantes, France}

\author{O.~V.~Rogachevsky}
\email{rogachevsky@jinr.ru}
\affiliation {Joint Institute for Nuclear Research (JINR), Dubna, Russia}

\author{H.~H.~Wolter}
\email{hermann.wolter@lmu.de}
\affiliation {Fakult\"at f\"ur Physik, Universit\"at M\"unchen, M\"unchen, Germany}

\date{\today}

\begin{abstract}
	We estimate the chemical freeze-out of light nuclear clusters for NICA energies of above 2 A GeV. 
	 On the one hand we use results from the low energy domain of about 35 A MeV, where medium effects have been shown to be important to explain experimental results. On the high energy side of LHC energies the statistical model without medium effects has provided results for the chemical freeze-out. The two approaches extrapolated to NICA energies show a discrepancy that can be attributed to medium effects and that for the deuteron/proton ratio amounts to a factor of about three. These findings underline the importance of a detailed investigation of light cluster production at NICA energies. 
\end{abstract}


\maketitle

\section{Introduction}
\label{Sec:Introduction}

The particle production measured in heavy ion collisions (HIC) is of interest to infer the properties of dense matter. The description of the time evolution of the fireball produced in HIC demands a non-equilibrium approach to describe the time dependence of the distribution function of the observed products, which are mainly neutrons, protons, and clusters at low energies, but also mesons, hyperons and antiparticles at high energies. Different transport codes have been developed to describe the time evolution of the fireball, but the formation of bound states (clusters) remains an open problem and semi-empirical assumptions such as the coalescence model have been applied.

In contrast, the composition of dense matter is well investigated in equilibrium. The chemical freeze-out concept assumes that 
the system is approximately in equilibrium as long as collisions are sufficiently frequent to establish the corresponding distributions. For an expanding fireball this is no longer the case at a critical density, so that the chemical equilibrium freezes out at the corresponding parameter values for temperature $T$, baryon number density $n_B$, and proton fraction $Y_p$. This concept has been proven to be an appropriate starting point to describe HIC at moderate laboratory energies $E_{\rm lab} = 35$ A\,MeV 
\cite{Kowalski:2006ju} that have been analyzed in this scheme in Ref.~\cite{Natowitz}, but also up to highest energies provided by heavy-ion collisions at the LHC \cite{Andronic,Randrup16}.

New facilities such as FAIR and NICA are under construction to investigate the region between these limiting domains, i.e. at temperatures of about 100 MeV and densities exceeding the saturation density. Of interest are the yields of particles and light clusters like neutrons (n), protons ({p}), deuterons ($^2$H, d), tritons ($^3$H, t), helions  ($^3$He, h), and $\alpha$-particles  ($^4$He). 
We discuss here what can be expected for this intermediate region, using a quantum statistical (QS) model, as discussed in Ref.~\cite{NICA}.

The freeze-out concept can only be considered as an approximation to describe disassembling matter. It has the advantage that correlations and bound state formation are correctly described within a QS approach. For a non-equilibrium theory, the equilibrium is a limiting case, and even more, the quasi-equilibrium (generalized Gibbs ensemble) serves to define the boundary conditions for the non-equilibrium evolution, see Ref.~\cite{Zubarev}.

As an intermediate step to full dynamical calculations the freeze-out concept has recently by combined with hydrodynamical calculations within the framework of the recently 
developed THESEUS event generator \cite{Batyuk:2016qmb} to produce more realistic 
freeze-out conditions.
The success of a local coalescence approach within the three-fluid-hydrodynamical (3FH) model \cite{Ivanov:2017nae}
in reproducting the rapidity distributions of light fragments measured by the  NA49 collaboration \cite{Anticic:2016ckv}
at SPS energies is encouraging. 
The local coalescence is in fact the same local thermal model 
where only the overall normalization is a free parameter.
The difference of this overall normalization from that predicted by 
the free-hadronic-gas model may indicate medium effects in the light-fragment production.

 \begin{figure}[t] 
	\includegraphics[width=0.5\textwidth,angle=-90]{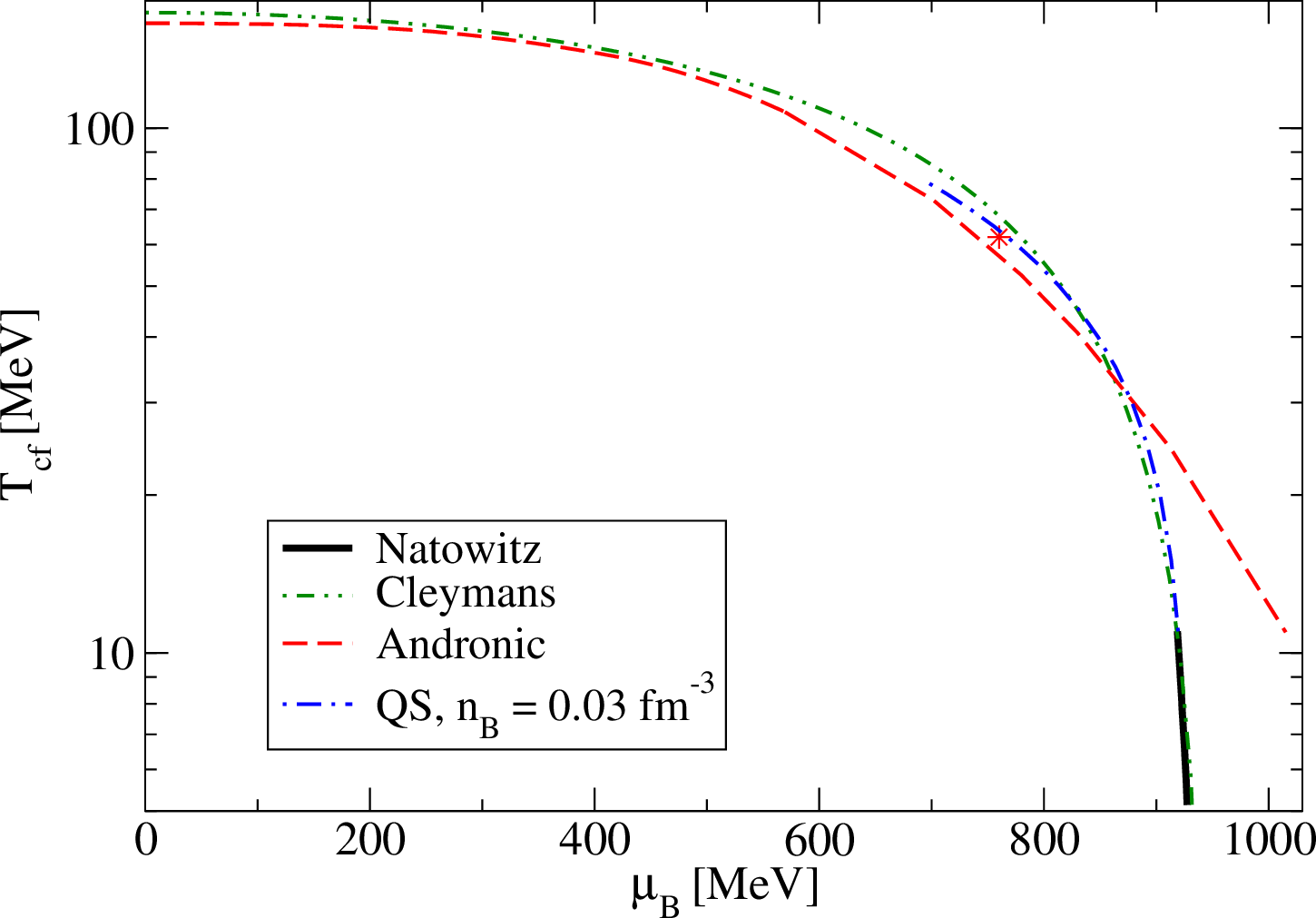}
	\caption{Freeze-out temperature as function of the baryon chemical potential. 
	The fit of Eq.(\ref{Eq:1}) according to Cleymans et al.~\cite{Cleymans} (green dash-double-dotted line), 
	is compared with the fit of Eq.(\ref{Eq:2}) of Andronic et al.~\cite{Andronic}  
	(red dashed line). 
	The star denotes the data point according to Ref.~\cite{Andronic} with lowest $T$. 
	Also shown is the freeze-out temperature derived from HIC at moderate energies \cite{Natowitz} (black solid line). 
	A calculation in the QS model with fixed freeze-out density $n_B=0.03$ fm$^{-3}$ (blue dash-dotted line) is also shown.}
 \label{Fig:1} 
 \end{figure} 
 
\section{Freeze-out parametrizations}

The freeze-out concept is surprisingly well appropriate to describe experiments at high energies~\cite{Cleymans}.
An empirical relation $T^{\rm Cleymans}_{cf}$ has been given to calculate it as function of the baryon chemical potential $\mu_B$,
\begin{equation}
\label{Eq:1}
\frac{T^{\rm Cleymans}_{cf}}{\rm GeV}=0.166-0.139 \left(\frac{\mu_B}{\rm GeV}\right)^2-0.053  \left(\frac{\mu_B}{\rm GeV}\right)^4
\end{equation}
with 
\begin{equation}
\frac{\mu_B}{\rm GeV}=\frac{1.308}{1+0.273 \sqrt{s_{NN}}/{\rm GeV}}
\end{equation}
and  $\sqrt{s_{NN}}=\sqrt{2 m_N E_{\rm lab}+2 m_N^2}$, $m_N=0.939$ GeV.
More recently in Ref.~\cite{Andronic}, using new data from LHC, another fit has been proposed:
\begin{equation}
\label{Eq:2}
T^{\rm Andronic}_{cf}=\frac{158.4 {\rm MeV}}{1+\exp[2.60-\ln(\sqrt{s_{NN}}/{\rm GeV})/0.45]},
\end{equation}
with 
\begin{equation} \mu_B=\frac{1307.5 {\rm MeV}}{1+0.288 \sqrt{s_{NN}}/{\rm GeV}}
\end{equation} 
The two parametrizations are compared in Fig.~\ref{Fig:1} and are seen to agree well for $T_{cf}> 60$ MeV. Also shown is a data point taken from Ref.~\cite{Andronic} with lowest temperature ($T \approx 62$ MeV, $\mu_B\approx 760$ MeV). At low temperatures, evidently, the fit of Andronic et al.~\cite{Andronic} is not applicable.
Freeze-out parameter values relevant for NICA energies are shown in Tab. \ref{Tab:1}.\\

\begin{table}[ht]
	\begin{tabular}{|c|c|c|c|c|c|}
		\hline
		$E_{\rm lab}$ [GeV] &$\sqrt{s_{NN}}$ [GeV]& $T^{\rm Cleymans}_{cf}$ [MeV] & $\mu^{\rm Cleymans}_B$ [MeV]&  $T^{\rm Andronic}_{cf}$ [MeV] & $\mu^{\rm Andronic}_B$ [MeV]   \\
		\hline
		2& 2.35 & 56.38 & 796.8 & 52.513 & 779.76\\
		3.85 & 3 &  79.956 & 719.07 & 72.93 & 701.45 \\
		9.84& 4.5 & 111.8 & 586.94 & 107.32 & 569.47 \\
		\hline
	\end{tabular}
	\caption{\label{Tab:1}
		Freeze-out parameter values relevant for NICA energies according to Cleymans et al. \cite{Cleymans} and Andronic et al. \cite{Andronic}.}
\end{table}

Freeze-out conditions are also well investigated experimentally at rather low laboratory energies, e.g., by  Natowitz et al.~\cite{Natowitz}.
The time evolution of the expanding fireball is deduced from the velocity of the emitted particles together with coalescence models. 
As seen in Fig. \ref{Fig:1}, the fit $T^{\rm Cleymans}_{cf}$, Eq. (\ref{Eq:1}), meets nicely the low-temperature data of Ref.~\cite{Natowitz}.
However, the law of mass action (LMA) \cite{DasGupta:1981xx}, also denoted as nuclear statistical equilibrium (NSE), was found to be not sufficient to explain the data, and medium effects have to be considered.
This was achieved  in the QS approach, described in the next section.
Alternatively, an excluded volume concept \cite{Hempel} has been used to include medium effects in a semi-empirical way.
An attempt to reproduce the parametrized chemical freeze-out line in the QCD Phase diagram from a kinetic condition has been made in Ref.~\cite{Blaschke:2017lvd}, involving chiral symmetry restoration and deconfinement. 

\section{Freeze-out densities}

 \begin{figure}[t] 
	\includegraphics[width=0.5\textwidth,angle=-90]{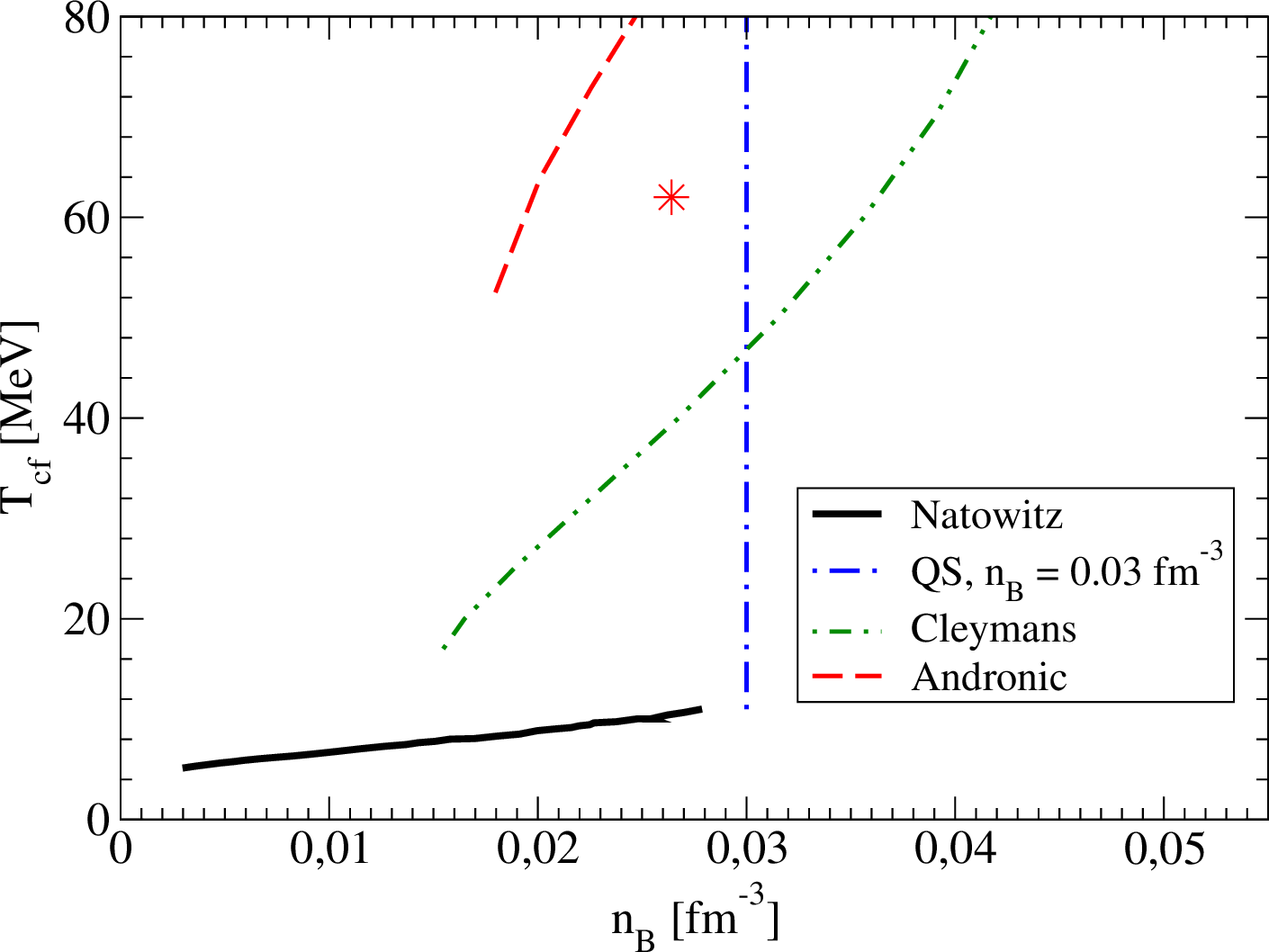}
	\caption{Freeze-out temperature as function of the baryon number density. Notations as in Fig. \ref{Fig:1}.}
 \label{Fig:2} 
 \end{figure}  

The determination of the baryon freeze-out density is not simple, as already known from HIC at moderate energies. 
In contrast to the freeze-out temperature which is well described by the yield ratios of different emitted particles, 
the freeze-out density is very sensitive to the chemical potential and the considered approximation. 

Th treat this problem a quantum statistical (QS) approach including light clusters \cite{R} is used here, which includes medium effects of nucleons and light clusters due to Pauli blocking and self energy shifts~\cite{R,R2}. 
The single-nucleon quasiparticle shift was taken according to the density-dependent relativistic mean-field approach (DD2-RMF) of Typel \cite{Typel}.
The nucleons $i = n, p$ with rest masses $m_i$ are treated as quasiparticles of energy
$E_i( p)=\sqrt{p^2 +(m_i-S_i)^2}+V_i$, with scalar and vector potentials, $S_i$ and $V_i$, which depend on density, proton/neutron asymmetry and temperature. A low density expansion of these potentials, which is useful in the present context, is given in the appendix.

The freeze-out densities at low temperatures are determined by the liquid-gas phase transition. 
Above the critical temperature of the liquid-gas phase transition of about 12 MeV the freeze-out density seems to remain nearly constant. A value $n_B \approx 0.03$ fm$^{-3}$ seems to be reasonable. From Ref.~\cite{Randrup16} this can be considered to be representative for the freeze-out at the NICA energies. But, as shown there, the freeze-out realistically dependes on density and temperature.

Using different expressions for the baryonic chemical potential at given temperature, the freeze-out temparature as a function the baryon density is shown in Fig. \ref{Fig:2}.
Also the result of Natowitz et al. \cite{Natowitz} is given. 

Here, dynamical transport simulations would be helpful to support the interpretation of the data with statistical models. The calculations using the coalescence model are of interest as also used in \cite{Natowitz}. Also the treatment with the the combined hydrodynamical and statistical model~\cite{Ivanov:2017nae} will give further insight.

\begin{figure}[t] 
	\includegraphics[width=0.5\textwidth,angle=-90]{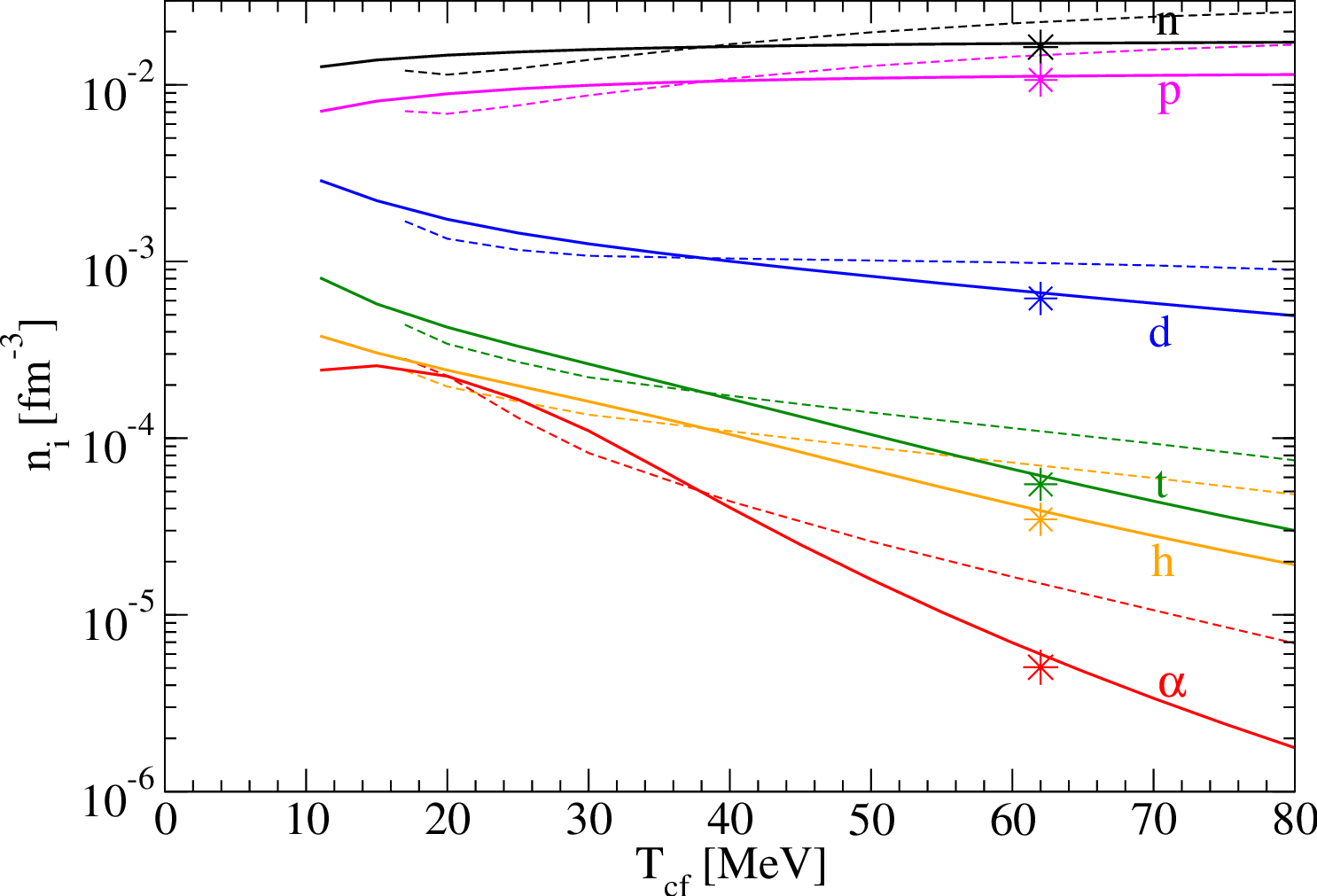}
	\caption{Densities of different constituents as a function of the freeze-out temperature calculated in the QS model for different choices of the chemical potential. Dashed lines: chemical potential according to Cleymans et al. \cite{Cleymans}, stars according to the lowest temperature value given by Andronic et al. \cite{Andronic}. Full lines: chemical potential from the QS model at fixed baryon density $n_B=0.03$ fm$^{-3}$.}
 \label{Fig:3} 
 \end{figure}  
 
 \section{Composition}
 
 The composition of the matter at freeze-out is shown in Fig. \ref{Fig:3} as calculated in the QS approach with different choices of the chemical potential. It is seen that the constant-density QS calculations matches fairly well with chemical potential taken from the parametrization of Cleymans et al.~\cite{Cleymans}, as one could have already expected from the results shown in Fig.~\ref{Fig:1}.
 This should also match with the low-energy results of Natowitz et al.~\cite{Natowitz}.

\begin{figure}[t] 
	\includegraphics[width=0.5\textwidth,angle=-90]{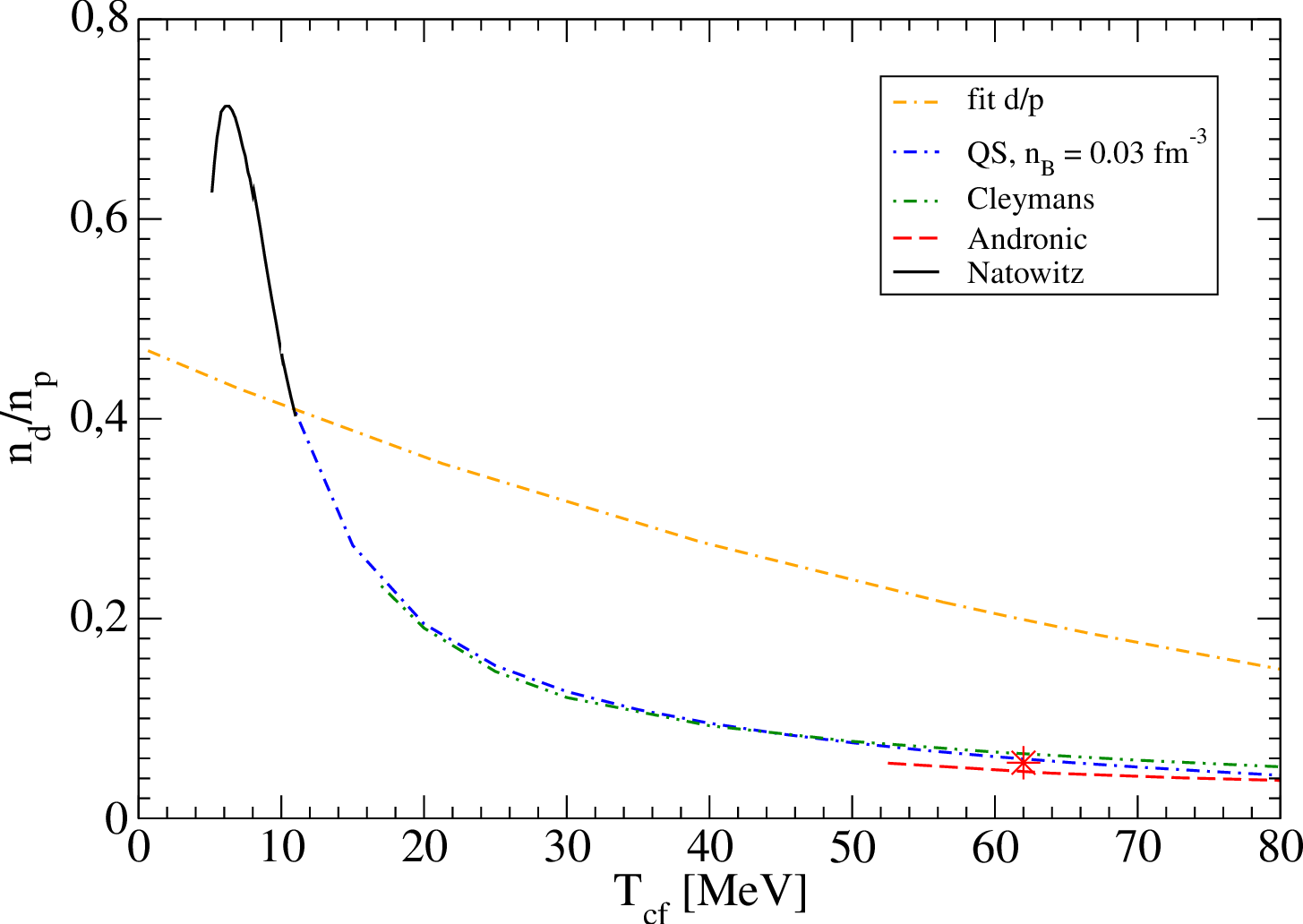}
	\caption{Deuteron to proton fraction at freeze-out temperature. Shown is the fit of Eq.\ref{Eq:4} (orange, dot-double-dash). The meaning and notations of the other curves are as in Fig. \ref{Fig:1}.}
 \label{Fig:4} 
 \end{figure}

 The d/p ratio is shown in Fig. \ref{Fig:4}. A fit formula \cite{Feckova:2015qza}
 \begin{equation}
d/p=0.8 [\sqrt{s_{NN}}/{\rm GeV}]^{-1.55}+0.0036
\label{Eq:4}
\end{equation}
which describes the d/p ratio well for energies above $\sqrt{s_{NN}}\sim 4 {\rm GeV}$ 
gives large values at the higher freeze-out temperatures of this figure.
In contrast, in the QS model we obtain a rather low value for the d/p ratio compared with the fit for all choices of the chemical potential. Reasons are the use of the full second virial coefficient and the Pauli blocking~\cite{R2}.
The QS calculation indeed matches well with the low-energy data of Ref.~\cite{Natowitz}.

\section{Conclusions}

HICs are non-equilibrium processes. The yields of particles and clusters should ideally be described by a transport approach which
includes also the formation, propagation, and collisions of bound states. The freeze out approach may be considered as a step in this direction,
by including the many-body correlations in a correct way. It describes the situation where relaxation to equilibrium is fast compared with the
time evolution of the thermodynamical parameters. This way, it serves a an ingredient (source term in the sense of the Zubarev approach)
to solve the dynamical evolution of the non-equilibrium system.

The results obtained from the freeze-out concept are surprisingly good. At high energies as well as at low energies,
the applicability of this concept has been demonstated. Here we are interested to combine both limiting cases to
obtain results for the intermediate region. In particular, the NICA facility is appropriate to investigate this region.

The density is very sensitive to the
value of the chemical potential, and the determination of the freeze-out density is a issue of future discussions.
Similarly, the composition is also very sensitive to the thermodynamic parameters as well as the treatment of medium effects.
Whereas the role of in-medium effects is clearly shown for freeze-out densities at low temperatures, the influence of
in-medium effects at intermediate temperatures such as relevant for the NICA experiments is under discussion.
This concerns in particular the ratio of deuterons to protons. We point out that this may be an important issue of future experiments at NICA.
We also plan to address this problem within the framework of recently 
developed THESEUS event generator \cite{Batyuk:2016qmb} that provides more realistic 
freeze-out conditions.

\subsection*{Acknowledgements}
This work was supported by the Russian Science Foundation grant No. 17-12-01427.

\subsection{Appendix: Low-density expansion}

A density dependent RMF model was considered in Ref.~\cite{Typel}. The following low-density expansions are derived from this model and reproduce the DD-RMF results below 
the baryon density $n\le0.2$ fm$^{-3}$ within 0.1 \%. Variables are the total baryon density 
$n=n^{tot}_n+n^{tot}_p$ in units of fm$^{-3}$, the asymmetry $\delta=(n^{tot}_n-n^{tot}_p)/n$, and the temperature $T$ in MeV. These expressions update the expansions given in the appendix of Ref.~\cite{R+}.
For the scalar part of the DD2-RMF we use the fit 
\begin{eqnarray}
S_i(T, n_B, Y_p) &=& (4463.117- 6.609841~T - 0.170252 \delta^2 + 4.111559 \delta^4) n_B
\times \frac{1+c_1 n_b+ c_2 n_B^2}{1+c_3 n_b+ c_4 n_B^2},
\nonumber \\ 
c_1&=&20.56456- 0.040985~T -0.339370~\delta^2 - 0.997156~\delta^4~,
\nonumber \\ 
c_2&=& 15.98022 + 0.866352~T - 2.020097~\delta^2 - 3.018041~ \delta^4~,
\nonumber \\ 
c_3&=& 24.27416 - 0.074176~T - 0.542662~\delta^2 + 1.196491 ~\delta^4~,
\nonumber \\ 
c_4&=& 114.5972 + 1.3497461~T + 2.674353~\delta^2 + 0.726793~\delta^4~,
\end{eqnarray}
where $\delta=(1-2Y_p)$, and $i=n,p$. For the vector part one obtains 
\begin{eqnarray}
V_p(T, n_B, Y_p)&=&(3403.144 + 0.000052~T- 486.581687~\delta - 2.420361~\delta^2) n_B 
\times \frac{ 1+d_1 n_b+ d_2 n_B^2}{1+d_3 n_b+ d_4 n_B^2},
\nonumber \\ 
	d_1&=&0.662946 - 0.006142~T- 1.140795~\delta - 0.717645~\delta^2,
	\nonumber\\
	d_2&=&10.77796 + 0.004432~T+ 0.80204~\delta + 0.457561~\delta^2,
\nonumber\\
	d_3&=&3.432703 + 0.000104~T - 1.548693~\delta - 0.336038~\delta^2,
\nonumber\\
	d_4&=&23.0145 - 0.033018~T - 5.922645~\delta + 0.050892~\delta^2 ,
\end{eqnarray}
\normalsize
and  $V_n(T, n_B, Y_p)=V_p(T, n_B, (1-Y_p))$.

\end{document}